# Exciton diffusion in amorphous organic semiconductors: reducing simulation overheads with machine learning

Chayanit Wechwithayakhlung[1,2], Geoffrey R. Weal[3,4,2], Yu Kaneko[5], Paul A. Hume[3,4], Justin M. Hodgkiss[3,4], Daniel M. Packwood[1,2*]

[1] Institute for Integrated Cell-Material Sciences (iCeMS), Kyoto University, Kyoto, Japan

[2] Center for Integrated Data-Material Sciences (iDM), MacDiarmid Institute for Advanced Materials and Nanotechnology, Wellington, New Zealand

[3] MacDiarmid Institute for Advanced Materials and Nanotechnology, Wellington, New Zealand

[4] School of Chemical and Physical Sciences, Victoria University of Wellington, Wellington, New Zealand

[5] Daicel Corporate Research Center, Innovation Park (iPark), Daicel Corporation, Himeiji, Japan

* Corresponding author. Email: dpackwood@icems.kyoto-u.ac.jp

## Abstract

Simulations of exciton and charge hopping in amorphous organic materials involve numerous physical parameters. Each of these parameters must be computed from costly *ab initio* calculations before the simulation can commence, resulting in a significant computational overhead for studying exciton diffusion, especially in large and complex material datasets. While the idea of using machine learning to quickly predict these parameters has been explored previously, typical machine learning models require long training times which ultimately contribute to simulation overheads. In this paper, we present a new machine learning architecture for building predictive models for intermolecular exciton coupling parameters. Our architecture is designed in such a way that the total training time is reduced compared to ordinary Gaussian process regression or kernel ridge regression models. Based on this architecture, we build a predictive model and use it to estimate the coupling parameters which enter into an exciton hopping simulation in amorphous pentacene. We show that this hopping simulation is able to achieve excellent predictions for exciton diffusion tensor elements and other properties as compared to a simulation using coupling parameters computed entirely from density functional theory. This result, along with the short training times afforded by our architecture, therefore shows how machine learning can be used to reduce the high computational overheads associated with exciton and charge diffusion simulations in amorphous organic materials.

# 1. Introduction

Efficient methods for simulating charge and energy transport in solid-state organic materials will accelerate the development of novel organic electronics and energy conversion devices. In organic materials, large reorganization energies tend to localize electronic states, resulting in a transport mechanism in which charge or energy carriers hop between molecules [1, 2]. Several methods for simulating carrier hopping exist, each of which involve physical parameters such as electronic couplings that need to be estimated beforehand *via* computationally intensive *ab initio* calculations. For the case of amorphous organic materials, a considerable number of *ab initio* calculations are typically required, as physical parameters need to be computed for each of the many unique intermolecular interactions and local environments which exist in the disordered system. This large number of calculations represents a huge computational overhead which must be surmounted before charge or energy dynamics can be simulated in an amorphous organic materials.

Excitons - electron-hole charge pairs that are bound together by Coulombic attraction - are important energy carriers in solid-state organic materials [3, 4, 5]. The diffusion of excitons between molecules is central to the function of organic semiconductor technologies, including photovoltaics (solar cells) [3, 6, 7], light-emitting diodes (LEDs) [8, 9, 10], lasers [11, 12], and photocatalysts [13 - 17]. For these technologies, the exciton diffusion rate is an important figure-of-merit for assessing candidate materials. Indeed, for the case of organic photovoltaics, high exciton diffusion rates are needed to ensure that the exciton can reach the heterojunction before the exciton decays *via* electron-hole recombination and other processes [3]. This effect in turn constrains the sizes of the donor and acceptor domains. Similarly, hyperfluorescent LEDs rely on efficient exciton diffusion in order to selectively funnel excitons to terminal emitter molecules [18, 19]. While exciton diffusion rates obtained from hopping-type simulations have achieved good agreement with spectroscopic measurements [20 - 22], the computational overheads of these simulations must be addressed before they can facilitate the large-scale screening of materials for technological applications.

Over the last several years, machine learning has emerged as a powerful means of reducing the demands of simulations in computational materials science [23, 24]. With machine learning, one can bypass the resource-intensive parts of a simulation with regression models trained on structure-property databases. For the case of exciton diffusion in organic materials, one might imagine substituting machine-learned regression models for the heavy *ab initio* calculations used to estimate exciton couplings or other physical parameters which enter into a hopping model. These physical parameters, which usually would be calculated one-by-one from first-principles, would instead be obtained with negligible computational cost by feeding them through a simple regression model.

Several groups have recently presented regression models for estimating intermolecular coupling energies in a variety of organic materials as well as in related biological systems. Lederer *et al* presented a kernel ridge regression (KRR) model trained to predict coupling parameters for charge hopping in pentacene, and used the model predictions in a subsequent kinetic Monte Carlo charge hopping simulation [25]. Similar works were subsequently reported by Wang *et al* [26], Kramer *et al* [27], Farahvash *et al* [28], and Gagliardi *et al* [29], which further demonstrated the ability of KRR-based models (and their Bayesian generalizations, Gaussian process regression (GPR) models) for quickly predicting coupling energy parameters in pentacene and other simple organic solids. In addition to KRR or GPR-based models, neural network models for predicting such coupling parameters have been presented by Farahvash *et al* [28], Wang [30], and Li *et al* [31]. Related works for the case of biological systems have been reported by Maiti and co-workers, in which neural network models were trained to predict coupling parameters between DNA bases [32, 33], and by Cignoni *et al*, who presented a KRR model for predicting exciton coupling energies between pigment molecules in a protein complex [34]. The high prediction accuracies achieved by these models, as well as their minuscule prediction times compared to *ab initio* calculations, point towards an optimistic future in which exciton transport simulations could be performed with relatively small computational overhead.

However, despite the success of the above studies, there is one important contribution to the computational overhead of an exciton hopping simulation which has not been addressed so far: the large training times required for typical machine learning models. Indeed, for models based on GPR or KRR – two of the most popular methods in computational materials science – model training times scale as $N^\beta$, where $N$ is the size of the training data set. Most of the studies cited above used enormous amounts of training data, ranging from a few thousand to a few hundred thousand instances, which necessarily implies enormous training times. Large training times add to the computational overhead of subsequent exciton diffusion simulations, reducing the overall advantage of using machine-learned models over *ab initio* calculations.

In this paper, we present a new machine-learning architecture for predicting exciton coupling parameters in amorphous organic semiconductors. For the training data sizes considered here ($N \sim 3000$), our architecture allows one to build regression models within around 25 % of the time required to build a standard GPR model. On the basis of a new type of sensitivity analysis, we confirm that our constructed model can be interpreted in terms of straightforward and consistent structural information, thereby supporting its generality beyond the training data. Moreover, we show that this model is sufficiently accurate for use in subsequent exciton diffusion simulations, in spite of the reduced training times. We apply our architecture for the case of exciton transport in amorphous pentacene, however it is not restricted to this particular material. This architecture can also be applied to charge transport in organic materials. This work therefore provides a means to reduce the computational overheads associated with hopping simulations of exciton and charge dynamics.

This paper is organized as follows. In section 2 we describe our methods and our machine-learning architecture for building predictive models of intermolecular exciton coupling. Section 3 describes our results, including the outcome of our model sensitivity analysis and the validation of our predicted coupling parameters in an exciton diffusion simulation. Discussion and conclusions are left to section 4.

## 2. Method

### 2.1. Generation of amorphous pentacene system

We study exciton diffusion in the amorphous pentacene system shown in Figure 1A. The amorphous pentacene configuration was generated by consecutive molecular dynamics simulations as follows. First, 800 pentacene molecules were placed at random into a 100 x 100 x 100 Å box using the Packmol package [35]. An amorphous phase was then generated in multiple steps. In the first step, a geometry optimization was performed. In the second step, velocities were assigned to each atom by sampling from a Boltzmann distribution at 600 K, and a sequence of 100 ps-long molecular dynamics simulations were performed at 1000 atm and temperatures of 500, 400, 600, 700, 600, 500, 600, 600, 250, and 300 K, respectively. Another simulation at 1000 atm and 300 K was then performed for 500 ps in order to induce molecule agglomeration. In the third step, a sequence of 2 ns-long simulations were performed at 100 atm at temperatures of 300 K, 250 K, 300 K, and 200 K in order to relax the molecule geometries and local environments around each molecule. In the fourth step, two 2 ns-long simulations were performed at 1 atm and temperatures of 200 K and 250 K, respectively, in order to induce the formation of amorphous pentacene. In the final step, the velocity vectors for each molecule were examined to confirm the absence of high-energy molecules. Each molecular dynamics simulation was performed using 2 fs time steps, and the simulation box was allowed to relax in each case. Simulations were performed using the GAFF2 force field [36] and the GROMACS package [37]. Periodic boundary conditions were applied throughout the entire procedure. Temperature and pressure regulation were applied using the Nose-Hoover thermostat [38, 39] and Berendsen algorithm [40], respectively. The amorphous pentacene configuration can be downloaded in CIF format in Supporting Information 1.

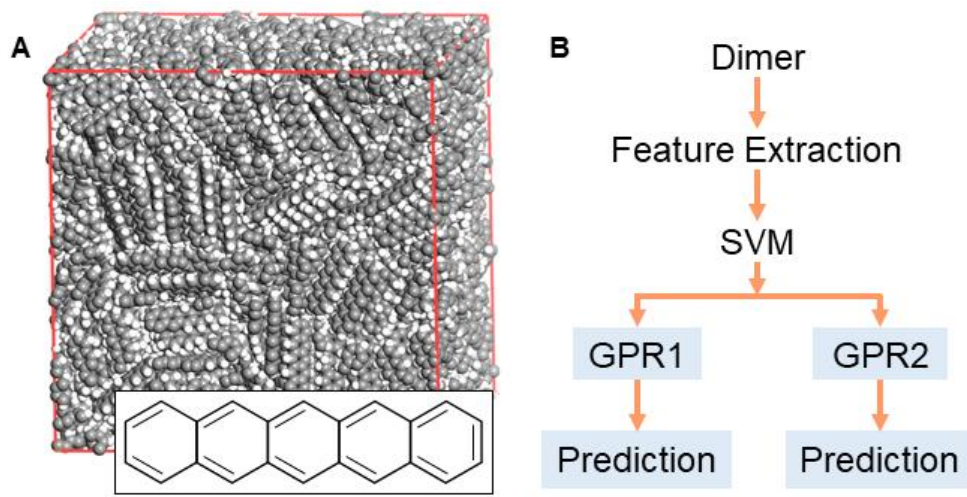

**Figure 1.** (A) Amorphous pentacene system considered in this study. Grey and white spheres correspond to carbon and hydrogen atoms, respectively. The chemical structure of pentacene is shown in the insert. (B) Summary of the machine-learning architecture for predicting exciton coupling energies for pentacene dimers. SVM indicates the support vector machine component. GPR1 and GPR2 indicate the two Gaussian process regression components.

Amorphous pentacene was selected as a model material because it is a representative small molecule semiconductor with a flat aromatic structure typical of many high-mobility cases. Pentacene also lacks the structural degrees of freedom arising from electronically inert side chains, which facilitates the generation of realistic amorphous packing structures. However, while our exciton coupling parameters and diffusion simulations consider singlet excitons, in real thin films of crystalline pentacene triplet excitons also play an important role when singlet excitons undergo singlet fission at particular sites [41]. For the case of amorphous pentacene, singlet exciton fission rates should be reduced due to the increased difficulty of reaching the single fission sites compared to the crystalline case. Nonetheless, we emphasize that the amorphous pentacene model used for these calculations is an ideal model system to test our machine-learning architecture, because it embodies a situation in which thousands of electronic configurations emerge from amorphous packing of a relatively simple molecular building block.

*2.2. Dimer extraction and density functional theory calculations*

Pentacene dimers were extracted from the amorphous pentacene system using an in-house dimer extraction code. A pair of molecules was considered a dimer if any two carbon atoms in the respective molecules were within 5.0 Å of each other. Dimers involving molecules spanning across the cell boundary (due to periodic boundary conditions) were included. In total, 4927 pentacene dimers were obtained. Structure files for these dimers are provided in Supporting Information 2. Our extraction code was written in Python 3 and used the Atomic Simulation Environment, NetworkX, and Pymatgen packages [42 - 44].

For a dimer composed of pentacene molecules *a* and *b*, the exciton coupling value is defined as

$$v_{ab} = \langle \psi_b | H | \psi_a \rangle, \qquad (1)$$

where $\psi_a$ and $\psi_b$ denote states in which a singlet excitation is localized on molecules *a* and *b*, respectively, and *H* is the Hamiltonian operator for the dimer. For each dimer, the exciton coupling energy was calculated using the electronic energy transfer (EET) method as implemented in the Gaussian 16 software package [45]. In this method, the exciton coupling energy is computed in a perturbative way by contracting the converged transition densities of the isolated molecules *via* the linear response time-dependent density functional theory (TDDFT) Hamiltonian of the dimer [46, 47]. Single-point excited state energies were calculated using the CAM-B3LYP functional [48], along with the 6-311+G(2d,p) basis set. Integrals were evaluated with an ultrafine integration grid and the accuracy of two-electron integrals set to $10^{-12}$. Linear response TDDFT calculations for the lowest 10 excited states were performed within the Tamm–Dankoff approximation.

*2.3. Machine learning architecture*

A family of models for predicting exciton coupling values for molecular dimers was constructed within a common machine learning architecture. This architecture consists of the four components shown in Figure 1B, and each model differs in the settings (choice of kernel functions, values of numerical parameters, etc.) which specify each component. The four components are a feature extraction procedure, a support vector machine (SVM), and two Gaussian procession regression estimators (GPR1 and GPR2).

In short, this architecture allows for reduced training times as follows. Rather than constructing a single GPR component valid for all possible exciton coupling values, this architecture uses two GPR components which are respectively trained to make predictions in a 'weak' and a 'strong' coupling regime only. These two GPR components can therefore be built using a smaller amount of training data than a single GPR component which is valid for all cases. Moreover, because the time cost of building a GPR estimator scales cubically (as $N^3$, where $N$ is the size of the training data), training two GPR components with a small amount of training data can be less time consuming than training a single component with a large amount of data. The SVM component needs to be trained in addition to the two GPR components, however this can be performed relatively quickly as explained below.

In order to select a model for predicting exciton coupling values, a two-stage procedure was used. In the first stage, multiple models were built using a small set of training data and a simplified training procedure. The results of this stage were then used in the second stage, in which a final model was built using a larger amount of training data and a rigorous training procedure.

In the following we describe the four components in detail.

*Feature extraction.* Feature extraction refers to the generation of a real-valued vector (a feature vector) to describe a pentacene dimer. The feature extraction process in our architecture involves three steps. Let $x$ denote a pentacene dimer. In the first step, a Coulomb matrix $\mathbf{M}(x) = [M_{ij}(x)]_{n \times n}$, where $n = 72$ is the number of atoms in the dimer, is generated [49]. For a pair of atoms $i$ and $j$ contained in the same molecule, $M_{ij}(x)$ is set to 0. If $i$ and $j$ are contained in different molecules, then $M_{ij}(x)$ is set to $1/R_{ij}$, where $R_{ij}$ is the distance between $i$ and $j$. In the second step, a vector $\mathbf{W}(x)$ is computed from the matrix $\mathbf{M}(x)$. $\mathbf{W}(x)$ has length $n$, and is obtained by selecting the largest element of each row of $\mathbf{M}(x)$ and sorting them in descending order. We write this vector as

$$\mathbf{W}(x) = \left( W_1(x), W_2(x), \ldots, W_n(x) \right). \tag{2}$$

In the third step, principal component analysis is used to reduce the dimensions of $\mathbf{W}(x)$. The resulting vector

$$\mathbf{U}(x) = \left( U_1(x), U_2(x), \ldots, U_{n_d}(x) \right), \tag{3}$$

where $U_k(x)$ is the $k^{\text{th}}$ principal component and $n_d < n$, is the feature vector for dimer $x$.

In order to set the feature extraction component, only the parameter $n_d$ needs to be specified. The models in this work used either $n_d = 4$, 5, or 6. Larger values of $n_d$ are unjustified, as the principal component analysis showed that over 99 % of the variation in the data was accounted for by the first six principal components.

*SVM.* Support vector machines (SVM) are a type of classifier. In this work, they are used to classify dimers as exhibiting either weak or strong exciton coupling values. Using the predictions of the SVM, the set of extracted dimers can be divided into two sets exhibiting weak and strong coupling, respectively. GPR1 and GPR2 can then be trained using data sampled from the weak coupling and strong coupling sets, respectively.

Intuitively, a SVM performs classification by transforming the feature vector $\mathbf{U}(x)$ into a high-dimensional space equipped with a linear hyperplane. This hyperplane is oriented in such a way that dimers with weak and strong exciton coupling energies appear on respective sides of the plane. Given the hyperplane, the classification for a dimer $x$ can be predicted according to the formula

$$h(x) = \text{sign}\left( y_i + \sum_{k=1}^{N} \alpha_k y_k \left( K(x_k, x) - K(x_k, x_i) \right) \right), \tag{4}$$

where $y_i$ is the classification of the $i^{th}$ dimer in the training data (equal to -1 or +1 for weak and strong coupling cases, respectively), $\alpha_k$ is a coefficient which is set during the construction of the hyperplane, and $N$ is the size of the training data [50]. Note that equation (4) holds for any choice of $i$. The function $K$ is the so-called kernel function; in essence, the transformation of the feature vectors takes place through $K$.

The SVM component requires specification of four settings. The first is the value of the parameter $v_c$, which is the value which defines which exciton couplings are classified as weak and which are classified as strong. For a dimer $x$, the coupling energy $v_x$ is defined as weak if $|v_x| < v_c$ and as strong otherwise. For the models used in this work, $v_c$ was set to 0.005 eV, 0.0075 eV, or 0.0082 eV. The second setting is the choice of kernel function. For the models used in this work, linear, radial, third-order polynomial, and sigmoid kernel functions were used (see reference [50] for their definitions). The third setting is the value of the parameter used in the kernel function. Each of the kernel functions involved here used a single parameter. The fourth setting is the value of the penalty parameter for training points which locate on the wrong side the hyperplane.

Compared to the GPR components discussed next, the SVM component can be trained relatively quickly even with a large amount of training data. For a specific choice of settings, the time required for building the hyperplane scales as $N$, the size of the training data [51]. In practice, however, the values of kernel parameter and cost parameter often need to be optimized in order to obtain a hyperplane with sufficient accuracy.

*GPR1 and GPR2*. GPR1 and GPR2 are regression estimators which are trained to make predictions for weak and strong exciton couplings, respectively. GPR1 and GPR2 are trained using dimers which have been classified as weak- and strong-coupling cases, respectively, by the SVM.

GPR1 and GPR2 are both constructed using the Gaussian process regression (GPR) method. GPR involves constructing a probability density on the space of candidate values for $v_x$, where $v_x$ represents the exciton coupling for a dimer $x$ [52]. This probability density is given by

$$g(v_x) = \frac{1}{\sqrt{2\pi s_x^2}} \exp\left( -\frac{(v_x - \mu_x)^2}{2s_x^2} \right), \tag{5}$$

where $g(v_x)$ can be interpreted as the probability of the exciton coupling is equal to $v_x$ given the training data. This probability maximizes at $\mu_x$, which can be deduced using a so-called Bayesian procedure. Applying this procedure yields (see [53] for proof)

$$\mu_x = \Sigma_x \Sigma^{-1} \mathbf{V}. \tag{6}$$

In equation (6), $\Sigma = \lambda \mathbf{I}_{N'} + [\sigma_{ij}]_{N' \times N'}$, where $\lambda$ is a positive parameter, $\mathbf{I}_{N'}$ is an $N'$ x $N'$ identity matrix, $N'$ is the training data size and

$$\sigma_{ij} = C(x_i, x_j) \tag{7}$$

is the covariance between $x_i$ and $x_j$. Intuitively, $C(x_i, x_j)$ measures the structural similarity between training dimers $x_i$ and $x_j$. The term $\lambda \mathbf{I}_{N'}$ is included to ensure numerical stability when computing the inverse of $\Sigma$ and does not have a physical meaning. $\Sigma_x$ is a 1 x $N'$ row matrix of covariances between dimer $x$ and the dimers in the training data, i.e., $\Sigma_x = (C(x,x_1), C(x,x_1), …, C(x,x_{N'}))$. $\mathbf{V} = (v_1, v_2, … ,v_{N'})^T$ is a column matrix of exciton coupling values from the training data. $\mu_x$ in equation (6) can be interpreted as a weighted sum of the coupling values in the training data, where the weights reflect the degree of structural similarity between $x$ and the dimers in the training data. The variance $s_x^2$ in equation (5) is given by

$$s_x^2 = C(x,x) - \Sigma_x \Sigma^{-1} \Sigma_x^T. \tag{8}$$

For both GPR1 and GPR2, $\mu_x$ is used as the predictor for the exciton coupling $v_x$. For both GPR1 and GPR2 in all models, we set $C(x_i, x_j)$ equal to the squared exponential function:

$$C(x_i, x_j) = b_0 e^{-d_{ij}^2}, \tag{9}$$

where

$$d_{ij}^2 = \sum_{k=1}^{n_d} b_k \left( U_k(x_i) - U_k(x_j) \right)^2 \tag{10}$$

and the constants $b_0$, $b_1$, …, $b_{nd}$ are parameters (known as hyperparameters in this context) and $n_d$ is the feature dimension specified in the feature extraction component.

The GPR components are specified by setting the values of the parameters $\lambda$ and hyperparameters $b_0$, $b_1$, …, and $b_{nd}$. In order to obtain a GPR component with sufficient accuracy, the hyperparameters $b_0$, $b_1$, …, and $b_{nd}$ almost always need to be optimized in some way. The time required for a single step of the optimization process scales as $N'^3$, where $N'$ is the size of the training data [54]. This scaling arises from need to invert the covariance matrix when computing equation (6). Moreover, this optimization takes place in a space of dimension 5 to 7 (one dimension for each of the $n_d + 1$ hyperparameters) and can therefore require very many iterations before convergence is reached. In order to achieve short training times, it is important that GPR1 and GPR2 are trained using small sets of training data.

*Model selection.* The models developed during the first stage of selection were trained using a set of 100 pentacene dimers along with their DFT-calculated exciton coupling energies. The same training data set was used to train each component. In order to reduce training times and therefore compare a larger number of models, the kernel and penalty parameters for the SVM component were set to $1/n_d$ and 1, respectively, and were not optimized. For training the subsequent GPR1 and GPR2 components, these 100 dimers were split into two subsets according to the predictions of the SVM component. GPR1 (GPR2) was then built by using 80 % of the weak coupling (strong coupling) subset as training data and the remaining 20 % as testing data. Hyperparameters were optimized with respect to the mean-square error (MSE) of the predictions with respect to the testing data. A full list of models tested during the first stage of selection are listed in Supporting Information 3.

For the second stage of selection, we trained a single model based on the results obtained above. The SVM, GPR1, and GPR2 components were each built using an independently selected sample of 1000 dimers. For each component, 800 dimers were used for training and 200 used as testing data. SVM kernel and penalty parameters were optimized with respect to the SVM classification fail rate compared to a set of test data. This optimization was performed using a grid search. The hyperparameters of the GPR components were optimized as described above.

All calculations related to model selection were performed in the R environment using custom code [55]. The e1071 package was used to fit the SVM components [56]. The L-BFGS-B gradient optimizer as implemented in R was used to optimize GPR hyperparameters [57].

2.4. *Kinetic Monte Carlo (kMC)*

Kinetic Monte Carlo (kMC) is a method for simulating diffusion dynamics. This method treats the exciton diffusion process as a random hopping process over a network of molecules. In our kMC simulations, each molecule corresponds to one of the pentacene molecules from the amorphous pentacene system in Figure 1A. Hopping takes place within the dimers extracted above, and for each dimer the hopping rate constant is a pre-defined constant. The rate constant for hopping from molecule *a* to *b* was defined according to Marcus theory [58, 59]:

$$k_{ab} = \frac{2\pi}{\hbar} \frac{|v_{ab}|^2}{(4\pi\lambda_r k_B T)^{1/2}} \exp\left(-\frac{(\Delta E_{ab} + \lambda_r)^2}{4\pi\lambda_r k_B T}\right), \tag{11}$$

where $v_{ab}$ is the exciton coupling between molecules *a* and *b*, $\hbar$ is the reduced Planck constant, $k_B$ is the Boltzmann constant, $T$ is temperature, $\Delta E_{ab}$ is the energy difference between molecule *a* and molecule *b*, and $\lambda_r$ is the reorganization energy associated with excitation energy transfer. In most simulations of exciton diffusion $\Delta E_{ab}$ is treated as a random variable sampled from a Gaussian distribution with mean zero [60]. In this work, we set $\Delta E_{ab}$ to zero and neglect energetic disorder, thereby allowing us to focus on comparing simulations using DFT-calculated excitonic couplings with those predicted from our model. Likewise, $\lambda_r$ was kept constant for all molecules, i.e. the effect of different local environments on the reorganization energy is neglected. Note that $k_{ab}$ is only non-zero for dimers extracted from our dimer extraction code (section 2.2). Dimers not extracted from our code were considered far apart such that their coupling was negligible.

Our kMC simulations were initialized at time $t = 0$ by randomly placing the exciton on one of the 800 pentacene molecules from the amorphous pentacene system. Denoting this molecule as $a$, a residence time was calculated according to

$$\Delta \tau_a = -\frac{\ln z}{\sum_{d \sim a} k_{ad}}, \tag{12}$$

where $z$ is a random number sampled between 0 and 1 and $d \sim a$ denotes all molecules neighboring molecule $a$. The simulation time was advanced by $\Delta \tau_a$ and the exciton was shifted to a neighboring molecule $b$ with probability

$$p_{ab} = \frac{k_{ab}}{\sum_{d \sim a} k_{ad}}. \tag{13}$$

The above process was iterated for molecule $b$, and so on, until the end of the simulation.

Periodic boundary conditions were applied so that the amorphous pentacene system was repeated indefinitely in all directions. The reorganization energy of a pentacene molecule was obtained as [61]

$$\lambda_r = \left(E_g^* - E_e^*\right) + \left(E_g - E_e\right), \tag{14}$$

where $E^*$ and $E$ denote the energies of the excited state and ground state, respectively, and subscripts $g$ and $e$ denote optimized geometries in the ground and existed states, respectively. These energies were calculated using the CAM-B3LYP functional and 6-311+G(2d,p) basis set, resulting in a value of $\lambda_r = 340.6$ meV.

All simulations were run for a simulation time of 1 ns. kMC simulations were repeated 10,000 times independently in order to compute probability distributions and statistics related to exciton diffusion.

**3. Results**

*3.1. Comparison of candidate models (selection stage 1)*

The results of the first stage of our model selection procedure are summarized in Figure 2A - E. Each point corresponds to one distinct choice of settings for the Feature Extraction, SVM, and initial hyperparameters (before optimization) for the two GPR components (a full list of model parameter values is provided in Supporting Information 3). Each point is presented as a pair of adjacent squares, where the color of the left- (right-) hand square corresponds to the mean-square error (MSE) of the GPR1 (GPR2) component when tested against test data. The lower the value of the mean-square error (MSE) (i.e., the more blue the point in Figures 2A – E), the more accurately the GPR components can predict the value of excitonic coupling. The points are positioned according to the dissimilarity of the model settings using the t-distributed stochastic neighbor embedding (t-SNE) technique [62]. Here, the dissimilarity between models $i$ and $j$ is defined as

$$D_{ij} = \delta_{K_i,K_j} + \delta_{n_{d_i},n_{d_j}} + \delta_{v_{c_i},v_{c_j}} + \delta_{h_i,h_j}, \tag{15}$$

where $\delta_{rs}$ is the Kronecker delta, and $K_i$, $n_{di}$, $v_{ci}$, and $h_i$ refer to the SVM kernel type, feature dimension, coupling strength cut-off ($v_c$), and initial GPR hyperparameter values used for the gradient optimizer, respectively, for model $i$.

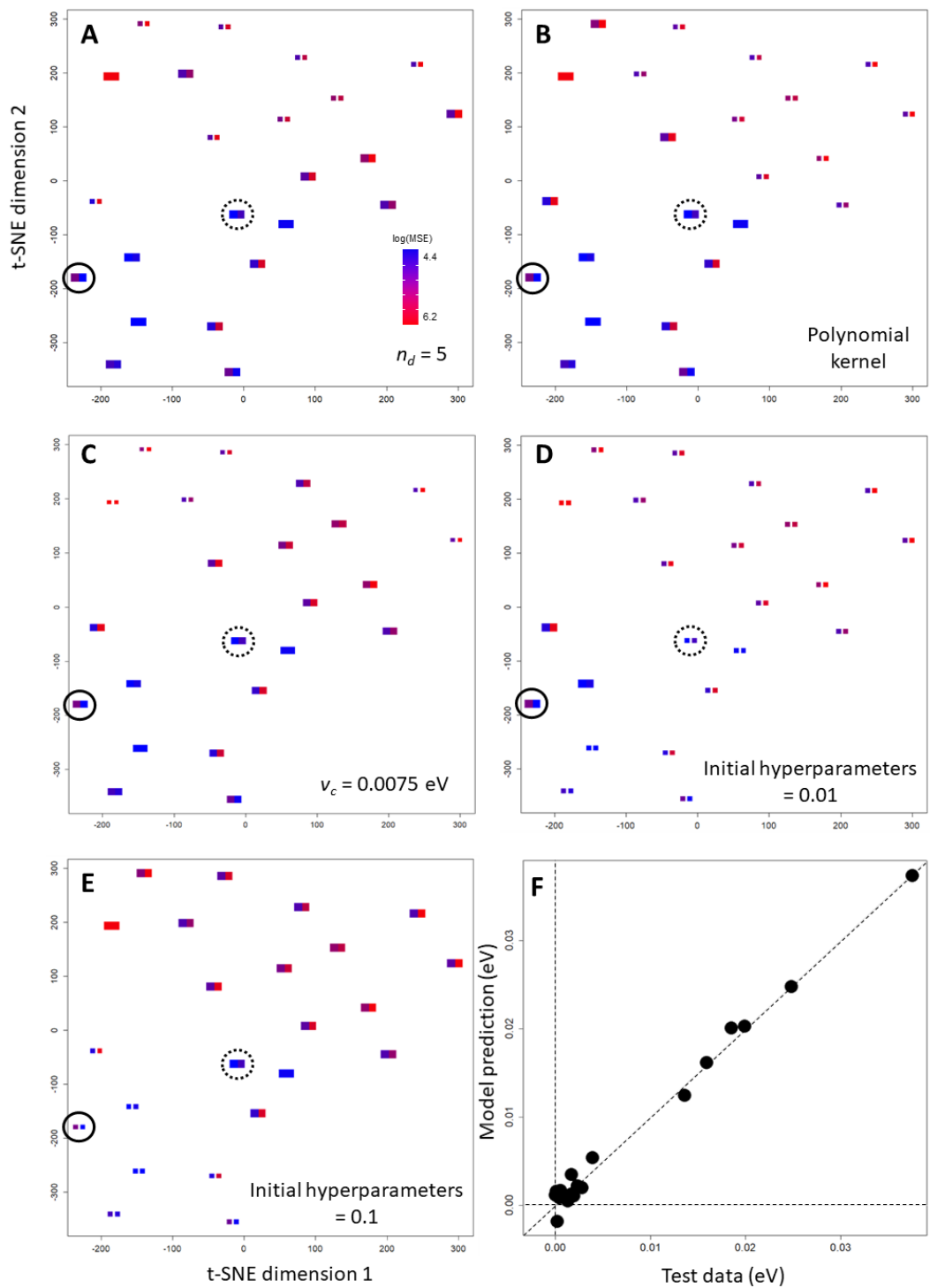

**Figure 2.** (A – E). Results of the first stage of model selection. Each point corresponds to one choice of settings for the Feature Extraction, SVM, and the two GPR components. The points are presented as adjacent squares, where the left (right)-hand square is colored according to the mean-square error (MSE) of GPR1 (GPR2) compared to test data. In each plot, we highlight the effect of certain settings by increasing the size of the corresponding points. The models which achieved the lowest MSE for GPR1 (GPR2) are indicated by the dotted (solid) circle (F) Performance of a selected model. This model used $n_d$ = 5, a polynomial kernel for the SVM component, $v_c$ = 0.0075 eV, and GPR1 (GPR2) covariance hyperparameters initialized at 0.1 (0.01). See text for details.

In Figures 2A – C, we can identify the common features of the Feature Extraction and SVM components of the low MSE models: $n_d = 5$, $v_c = 0.0075$ eV, and third-order polynomial kernels for the SVM component. Models using these settings can be identified by the enlarged points. For the final model, the Feature Extraction and SVM compounds should therefore be built using these settings.

For the purpose of building a final model, it is more useful to compare the initial values used for hyperparameter optimization rather than the final optimized hyperparameter values directly. This is because the hyperparameters will need to be optimized again when dealing with a new set of training data. According to Figures 2D – E, the hyperparameters for GPR1 and GPR2 should be initialized with different values in order to achieve good performance. In particular, the hyperparameters of GPR1 should be initialized at 0.1 and those of GPR2 should be initialized at 0.01 in order for the gradient optimizer to reach a good set of final values. In Figures 2A – E, two models that initialized GPR1 and GPR2 in this way are indicated by the dotted and solid circles. For these two models, we also find that $\lambda = 0.0001$ and 0.001, respectively. This indicates that the GPR components should also be built using different values of $\lambda$.

The results above therefore suggest that a good model could be built using the following settings: $n_d = 5$ for the Feature Extraction component; $v_c = 0.0075$ eV and third-order polynomial kernels for the SVM component; $\lambda = 0.0001$ and initial hyperparameter values of 0.1 for the GPR1 component; and $\lambda = 0.001$ and initial hyperparameter values of 0.01 for the GPR2 component. Figure 4F shows the predictive performance of such a model. The agreement between predicted and DFT-calculated exciton couplings is impressive, however due to the small size of the training and test sets, such agreement is unlikely to hold for all dimers in the amorphous pentacene system.

*3.2. Model interpretation using sensitivity analysis*

We now investigate whether the high-performing model in Figure 2C can be interpreted in a physically meaningful way. To this end we proceed in two steps.

In the first step, we plot the vectorized Coulomb matrices **W**($x$) (section 2.3 equation (2)) for each of the 100 dimers in the training data. This plot is shown in Figure 3A as a 100 x 72 matrix. The cell in row $i$ and column $r$ corresponds to $W_r(x_i)$, the $r^{th}$ element of the vector for dimer $x_i$. The cell is red if the Coulomb interaction in $V_r(x_i)$ is between two H atoms, gray if between an H and a C atom, and blue if between two C atoms. The left and right edges of the matrix in Figure 3A are clearly dominated by interactions involving hydrogen atoms. Indeed, interactions involving carbon atoms do not appear on the left-hand side of the plot at all until column $r = 10$. For the columns $r = 1$, 2, and 3, interactions between hydrogen atoms alone are observed for 64 %, 64 %, and 53 % of the dimers, respectively. Similarly, for column $r = 72$ on the right-hand edge of the plot, 11 % of the interactions are between carbon and hydrogen, and 89 % between pairs of hydrogen atoms.

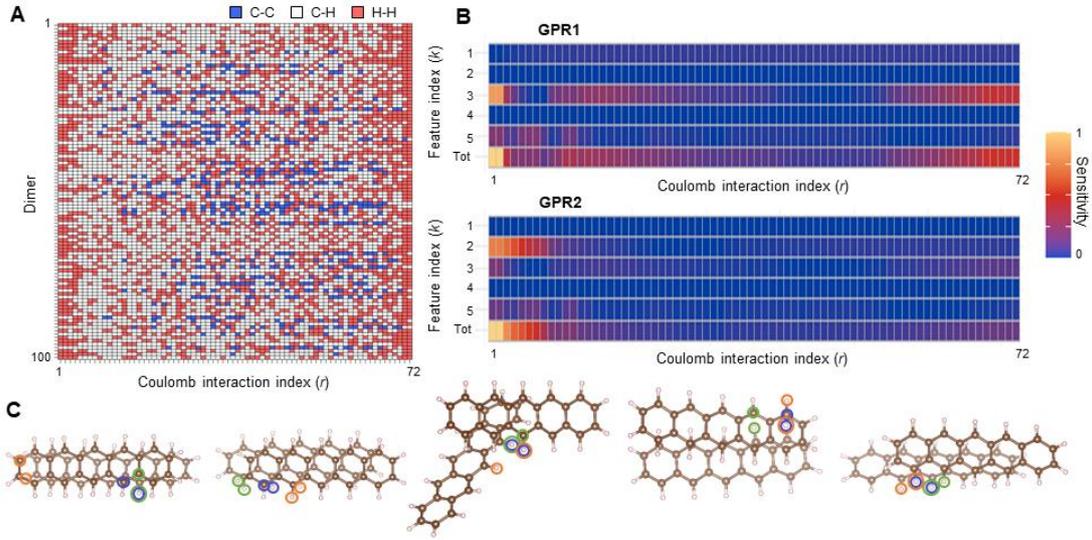

**Figure 3.** Sensitivity analysis of the selected high-performing model. (A) Plot of the sorted vectorized Coulomb matrices for the dimers in the training set. Rows correspond to one of the dimers, and columns correspond to the element of the vectorized Coulomb matrix. Colors correspond to the type of Coulomb interaction. (B) Plot of the sensitivities of each of the model input features. Rows correspond to input features and columns to the element of vectorized Coulomb matrix. The top plot is for the GPR1 component and the bottom one for the GPR2 component. Sensitivities are normalized by the maximum and minimum values in each table. See text for details. (C) The five dimers from our training set which exhibit the strongest exciton coupling. The pairs of atoms involved in the largest Coulomb interactions are indicated by the colored circles.

In the second step, we determine which Coulomb interactions GPR1 and GPR2 are most sensitive to. To do this, we expand the features $U_k(x_i)$ as

$$U_k(x_i) = \sum_{r=1}^{n} c_{kr} W_r(x_i), \tag{16}$$

where $c_{kr}$ are expansion coefficients. Equation (16) follows from the definition of principal components. Inserting equation (16) into equation (10) gives

$$d_{ij}^2 = \sum_{k=1}^{n_d} \frac{1}{b_k} \left( \sum_{r=1}^{n} a_{kr} \left( W_r(x_i) - W_r(x_j) \right) \right)^2, \tag{17}$$

where the constants $a_{kr} = b_k c_{kr}$ are referred to as *sensitivities.* Large $a_{kr}$ means that $d_{ij}$ is sensitive to the difference $W_r(x_i) - W_r(x_j)$. This sensitivity is transmitted to the predicted exciton coupling through the covariances (see equations (6) and (9)). In Figure 3B, we plot the sensitivities $a_{kr}$ as matrices for the components GPR1 (top) and GPR2 (bottom) as obtained from the high-performing model constructed at the end of the previous section. In both plots, the largest sensitivities tend to be found for small and large values of the index *r*. As shown in the previous paragraph, these values of *r* are mainly associated with hydrogen atoms.

The predictions of GPR1 and GPR2 are therefore sensitive to changes in the distances between hydrogen atoms of the two molecules. This shows that the high-performing model makes its predictions on the basis of a low-dimensional structural representation consisting of hydrogen atoms. These hydrogen atoms therefore act as a proxy for describing the alignment of the two pentacene molecules in the dimer. Figure 3C shows the five dimers from our training set that exhibit the strongest exciton coupling. The atoms involved in the three largest Coulomb interactions are indicated. Each of these interactions involves a hydrogen atom, which shows that hydrogen atom positions are important for predicting coupling in the strong-coupling regime.

The fact that this model has a straightforward physical interpretation suggests that its performance is based on a general structure-property relationship extracted from the training data. In other words, it suggests that the model accuracy is not based on 'overfitting' of the hyperparameters to the training data, and that the model can be meaningfully applied to dimers outside of the training data.

*3.3. Final model training times and performance (selection stage 2)*

Our final model for predicting exciton coupling was built using the settings shown in Supporting Information 3. These settings are identical to those of the high-performing models found from the first stage of selection, however the optimized values of the hyperparameters for GPR1 and GPR2 differ slightly due to the different training set used here. The parameters for the SVM component also differ slightly from the models used in the first stage of selection due to the use of the optimization procedure.

Training times were evaluated by running our script within the R console in a Kubuntu environment running on an Intel Xeon 3.5 GHz processor. This entire script includes the whole gamut of the calculation, from processing the structure files of the dimers, generating the feature vectors, optimizing the SVM parameters, and optimizing the hyperparameters of GPR1 and GPR2. The script runs these parts sequentially on a single processor. This script required 19.0 hours to complete its run. In fact, this training time could be reduced to around 9.5 hours if the script were parallelized so that GPR1 and GPR2 were trained on different processors, as the other parts of the script require negligible time to complete. In order to compare this training time, we considered the case of a single GPR model initialized with the same hyperparameter settings as GPR1 and using 2400 dimers for training and 600 for testing. This case therefore compares training times afforded by our machine learning architecture to those of an architecture consisting of a single GPR component, while holding the total amount of training data constant (recall that in the final model, the three components SVM, GPR1, and GPR2 were each trained using independently sampled sets of 800 training data points and 200 test data points). This script required 76.4 hours to complete its run. The training time of 19 hours for our final mode therefore corresponds to 25 % of the training time for the case of a single GPR model using the same amount of training data (or around 12.4 % if GPR1 and GPR2 were trained in parallel). This reduction can be traced to the fact that GPR training times scale as $N^\beta$, where $N$ is the size of the training data set.

When discussing training times, it is important to emphasize that GPR1 and GPR2 were implemented with a more general Gaussian process framework than the GPR models reported in previous papers. As is shown by equation (10), our GPR components contain a different hyperparameter per feature dimension. In contrast, previous works have used the more common GPR implementation in which equation (10) only uses one hyperparameter (that is, $b_1 = \cdots = b_d = b$) [25 - 29]. While our implementation of GPR is advantageous when trying to model complex functions such as exciton coupling, model training (hyperparameter optimization) necessarily takes places in a higher dimensional space. Our total training times are therefore not directly comparable to the training times of other GPR models reported so far. However, the $N^\beta$ scaling of the time required *per iteration* is independent of feature dimension, meaning that our observations of the faster training times afforded by our architecture will hold for other types of GPR implementations as well.

In Figure 4A, predictions of the final model are compared to the TDDFT-calculated couplings for all extracted pentacene dimers. While not perfect, the correlation between the predictions and DFT-calculation couplings is satisfactory: linear regression on the data in Figure 4A yields the result $y = (0.0024 \pm 0.0001) + (0.80 \pm 0.006)x$, where errors refer to one standard error, and $R^2 = 0.77$. In Supporting Information 4 we also report the results of a sensitivity analysis for the final model, which are similar to the ones reported for the model in the previous section. The final model is therefore also physically interpretable, suggesting that its predictions are based on a genuine structure-property relationship extracted from the training data.

In Supporting Information 5, we compare the predictions of the final model above with those of a single GPR model constructed using 800 dimers for training and 200 for testing. We are therefore comparing the accuracy achieved by our machine learning architecture with that of an architecture consisting of a single GPR component, while holding the total training time roughly constant. While both models achieve similar accuracy in the weak coupling regime, the single GPR model significantly underestimates coupling values in the strong coupling regime. This would be a serious concern if one wished to use such predictions in a subsequent exciton diffusion simulation, as the strongly coupled dimers are expected to have a decisive influence on diffusion dynamics. The inability of the single GPR model to make accurate predictions in the strongly coupled regime is unsurprising, because in a random sample of dimers weak coupling cases will be much more frequent than strong coupling cases, and the former will exert an oversized influence during model training. This problem is avoided by our architecture, which builds two GPR components trained specifically for the respective regimes.

*3.4. Kinetic Monte Carlo simulation*

The exciton coupling parameters predicted from the final model above were used as inputs in the kinetic Monte Carlo (kMC) simulation. For comparison, a 'benchmark' simulation using coupling parameters computed entirely from TDDFT was also performed. Figure 4B plots the mean-square displacement (MSD) of the exciton, as computed according to the formula

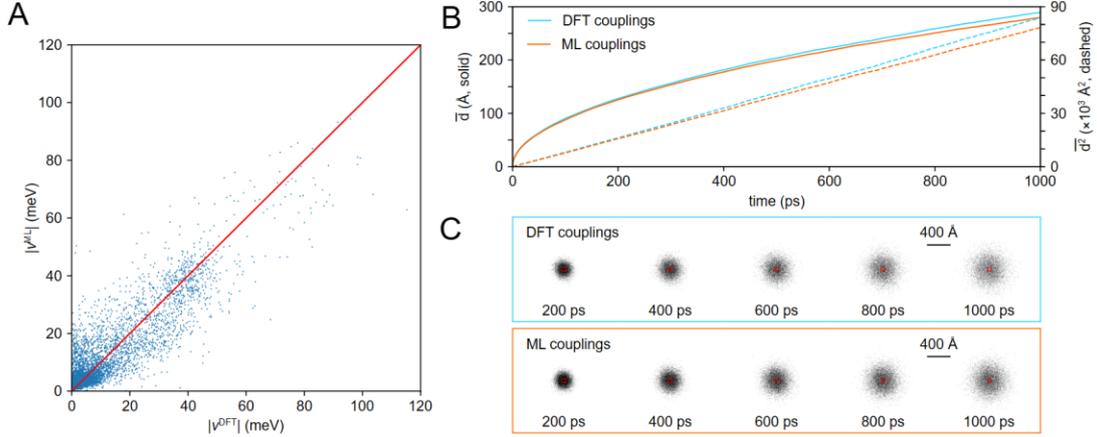

**Figure 4.** Final model performance and exciton diffusion simulations. (A) Predicted exciton coupling energies of the final model ($v^{ML}$) compared to *ab initio* calculations ($v^{DFT}$) for all extracted dimers. (B) Mean-square displacement of an exciton from its initial position as obtained from kinetic Monte Carlo simulations using model-predicted exciton coupling energies (orange lines) and *ab initio*-calculated couplings (blue lines). (C) Snapshots of the probability distribution of exciton positions at various times for the case of model-predicted and *ab initio*-predicted exciton couplings. For clarity, the probability distribution is plotted in the *xy* plane only. The red box indicates the dimensions of the simulation cell.

$$MSD(t) = \left\langle \left( \mathbf{r}_t - \mathbf{r}_0 \right)^2 \right\rangle \tag{18}$$

where $\mathbf{r}_t$ denotes the position of the exciton at time *t* and the angular brackets indicate averaging over the simulation runs. The mean-square displacement for the simulation performed using the predicted exciton couplings compares well to the benchmark simulation in both magnitude and time evolution. The exciton diffusion coefficient, which was computed according to

$$D = \frac{1}{6} \lim_{t \to \infty} \frac{MSD(t)}{t}, \tag{19}$$

was (1.547 ± 0.005) x 10$^{-3}$ cm$^2$ s$^{-1}$, which compares well to the value of (1.630 ± 0.02) x 10$^{-3}$ cm$^2$ s$^{-1}$ obtained from the benchmark calculation (see Table 1).

Figure 4C plots the probability distribution of the exciton position at different points in time, as projected onto the *xy* plane. Good agreement with the benchmark results is obtained: not only does the spread in the probability distributions look comparable, but the shape of the distribution is preserved. In order to compare the probability distributions in a quantitatively rigorous way, we calculate the eigenvalues of the diffusion tensor. These three eigenvalues correspond to the three dominant directions along which the probability distribution spreads, and their magnitudes correspond to the rate of spread in these directions. For the case of the simulations performed using predicted exciton couplings, these eigenvalues were computed to be (1.690 ± 0.020), (1.492 ± 0.007), and (1.460 ± 0.020) x 10$^{-3}$ cm$^2$ s$^{-1}$, respectively (see Table 1). These values compare to (1.820 ± 0.02), (1.550 ± 0.02), and (1.530 ± 0.02) x 10$^{-3}$ cm$^2$ s$^{-1}$, respectively, as obtained from the benchmark simulation. The eigenvalues for the case of predicted couplings are of a slightly smaller magnitude than those for the benchmark simulation, suggesting that the exciton coupling is underestimated by our model for the most strongly coupled dimers in the system. However, the ratios of the eigenvalues relative to the smallest one were quite similar (1.15:1.02:1.00 for the case of predicted couplings and 1.19:1.02:1.00 for the case of the benchmark), confirming that the simulation using predicted couplings correctly preserved the anisotropic nature of the exciton diffusion. The individual elements of the diffusion tensor are compared in Supporting Information 6, where a similar agreement with the benchmark case is observed: comparable but slightly smaller magnitudes, but very similar relative values as expressed by ratios.

|  | *Ab initio* couplings | Model-predicted couplings |
|---|---|---|
| Diffusion Coefficient ($\times 10^{-3}$ cm$^2$s$^{-1}$) | 1.630 ± 0.011 | 1.547 ± 0.005 |
| Diffusion tensor eigenvalues ($\times 10^{-3}$ cm$^2$s$^{-1}$) | | |
|   Major | 1.815 ± 0.014 | 1.686 ± 0.017 |
|   Middle | 1.551 ± 0.016 | 1.492 ± 0.007 |
|   Minor | 1.525 ± 0.012 | 1.462 ± 0.014 |
| Major: Middle: Minor | 1.19: 1.02: 1.00 | 1.15: 1.02: 1.00 |

**Table 1.** Exciton diffusion parameters estimated from kinetic Monte Carlo (kMC) simulations using *ab initio*-calculated and model-predicted exciton coupling energies. Error bounds correspond to one standard deviation.

In Supporting Information 6, we consider the case of a model built using a much larger training set of 2000 dimers for GPR1 and 2200 dimers for GPR2. kMC simulations using the predictions of this model achieved very similar results to the ones obtained from the model above. Interestingly, the magnitudes of the diffusion tensor eigenvalues and elements remain slightly underestimated for this case as well. We will return to this point in the next section. The agreement between the predictions of kMC using this model and the model described above suggests that the latter has converged, in some sense, with respect to training data size. The fact that this convergence can be reached even with small sets of training data may relate to the observations in section 3.2, which suggested that our models have a meaningful physical interpretation hence should generalize well beyond the small training and test data sets.

## 4. Discussion and conclusions

In order for machine learning to accelerate simulations for organic photovoltaics and other types of materials, it is important that the predictive models for simulation parameters can be built with short training times. Long training times may offset the reductions in computational time achieved through the final machine-learned model. In this paper, we presented a new machine learning architecture for predicting intermolecular exciton coupling values in amorphous organic solids. For the training set sizes used here (around 3000), our architecture allowed for predictive models to be trained within around 25 % of the time required to train a typical Gaussian process regression model (or, equivalently, a typical kernel ridge regression model). Importantly, when these predicted coupling energies were used as inputs in a subsequent exciton diffusion simulation, we obtained results which were in excellent agreement with a benchmark simulation, in spite of the reduced training time of the underlying model. Thus, for the purpose of predicting model parameters for subsequent exciton diffusion simulations, highly accurate machine-learned models built with massive sets of training data and long training times are not necessarily required. We partly attribute the accuracy of our model to the results of our sensitivity analysis, which suggest that the model can be interpreted in terms of hydrogen atom positions. This straightforward physical interpretation suggests that the model's predictions are based on a genuine structure-property relationship acquired from the data, rather than the result of over-training on the small training data set.

The accuracy of the model in spite of the reduced computational times might also be attributed to the special machine-learning architecture used to generate it. Instead of attempting the difficult task of fitting a single regression model valid for all possible molecule dimers, this architecture separates the possibilities into a 'weak coupling' and a 'strong coupling' regimes and builds regression estimators for these regimes separately. A similar strategy was used in the context of dimer interaction energies in reference [63]. While this model was built for the special but illustrative case of amorphous pentacene, the underlying machine-learning architecture is not restricted to this case and could be applied to other types of amorphous organic solids, including ones composed of large flexible molecules, or ones involving multiple molecular species.

While kinetic Monte Carlo (kMC) simulations using model-predicted exciton coupling parameters achieved excellent agreement with a benchmark simulation, the magnitudes of the diffusion tensor elements and eigenvalues were slightly underestimated. This underestimation was observed both for the model described above and for another model built using a much larger set of training data. For practical purposes such as comparing candidate materials for organic photovoltaics, such a shortcoming is not expected to be an issue. However, it suggests that multiple examples of dimers with very high exciton couplings might need to be included in the training data in order to obtain very accurate predictions of the diffusion tensor elements. On the one hand, these high-coupling cases may be more important in determining the magnitude of the diffusion tensor elements than the weak coupling cases. On the other hand, such cases are very rare, and might not have been included with sufficient frequency in the training sets used to build our models. These points should be clarified further in the next stage of this research. Concretely, additional theoretical work should be performed to determine what types of dimers are needed to obtain accurate predictions for the diffusion tensor elements, and novel sampling schemes for generating training sets of dimers developed accordingly. Theoretical work should also be performed to derive uncertainty bounds for diffusion tensor elements and other quantities in terms of training set sizes. The machine learning literature contains numerous uncertainty bounds for various types of regression and classification models [50], however these are expressed in terms of highly general concepts such as hypothesis space complexity and are difficult to apply to the case of exciton diffusion directly.

For applications in chemistry and physics, it is desirable that machine-learned models can be interpreted in terms of straightforward structural information. Such physical interpretations help clarify the structural motifs the model uses to make its predictions (in this case, relative positions of the hydrogen atoms in the two molecules of the dimer). In this paper, we employed a sensitivity analysis to probe for a physical interpretation of the GPR components in our models. This amounts to performing a *post-hoc* check on model after it is constructed. An alternative strategy to guarantee model interpretability is to design the regression model by deliberately including some aspects of the excitonic transfer mechanism and dimer atomic structure in the model's structure. This is difficult to do with the SVM or GPR models used in this paper, as much of the model's internal structure is hidden from view by the kernel and covariance functions. However, outside of the field of organic semiconductors, several groups have proposed neural network models which incorporate known physical relationships in their architectures (e.g., [64]). Compared to the sensitivity analysis approach, such models have some disadvantages. For example, the decision as to what physics should be incorporated into the model is subjective. Moreover, these models may acquire additional physical information during the training procedure. Ironically, such information could only be identified by performing a sensitivity analysis on the model following training. Finally, these models are more difficult to train due to the presence of additional parameters describing the incorporated physics. Indeed, if one insists that these additional parameters only take values within a physically sensible range, then it may become difficult to minimize model prediction errors sufficiently during training. Until these problems are overcome, *post-hoc* sensitivity analyses are probably the most practical means to ensuring that a machine-learned model is physically interpretable.

It is also important to consider the extent to which machine learning can be applied to simulate exciton diffusion in amorphous organic solids. In this work, a machine-learned model was substituted for the resource-heavy time-dependent density functional theory (TDDFT) method for computing exciton coupling parameters. However, these parameters are not the only ones present in a hopping simulation. For the case of the simulations performed here, which used a Marcus-type expression for the hopping rate, an energetic disorder parameter is also present. Moreover, some simulations include the so-called Huang-Rhys factor, which characterizes electron-phonon coupling strengths. Such electron-phonon coupling parameters need to be properly incorporated into simulations which describe exciton and charge dynamics as a function of temperature. Hence, there exists scope to reduce the computational demands of exciton diffusion simulations further by developing predictive models for these parameters on the basis of machine learning as well.

This work has introduced the strategy of splitting the training data into two groups in order to mitigate the unfavorable $N^\beta$ scaling of GPR or KRR-based models of exciton coupling. However, these kinds of models are widely used in materials informatics beyond simulations of charge or energy transport. Could this strategy therefore be used for materials discovery purposes, in which machine-learned models need to be fit to massive databases of candidate materials? We will explore this direction in future research.

**Acknowledgements**

This work has been supported by the Kyoto University On-Site Laboratory Initiative, the Institute for Integrated Cell-Material Sciences (iCeMS), the MacDiarmid Institute for Advanced Materials and Nanotechnology, and the Victoria University of Wellington high-performance computing cluster Rāpoi. GRW, PAH, and JMH acknowledge support from the Marsden Fund of New Zealand. DMP acknowledges JSPS Kakenhi Grant 21K05003.

# Supporting Information for:
# Exciton diffusion in amorphous organic semiconductors: reducing simulation overheads with machine learning

Chayanit Wechwithayakhlung[1,2], Geoffrey R. Weal[3,4,2], Yu Kaneko[5], Paul A. Hume[3,4], Justin M. Hodgkiss[3,4], Daniel M. Packwood[1,2*]

[1] Institute for Integrated Cell-Material Sciences (iCeMS), Kyoto University, Kyoto, Japan

[2] Center for Integrated Data-Material Sciences (iDM), MacDiarmid Institute for Advanced Materials and Nanotechnology, Wellington, New Zealand

[3] MacDiarmid Institute for Advanced Materials and Nanotechnology, Wellington, New Zealand

[4] School of Chemical and Physical Sciences, Victoria University of Wellington, Wellington, New Zealand

[5] Daicel Corporate Research Center, Innovation Park (iPark), Daicel Corporation, Himeiji, Japan

* Corresponding author. Email: dpackwood@icems.kyoto-u.ac.jp

**SI 1. Amorphous pentacene system**

CIF file of the amorphous pentacene system (Figure 1A)

**SI 2. Pentacene dimers**

Zip file of extracted pentacene dimers in xyz format.

**SI 3. Parameters of the tested models and parameters of the final model**

Excel file of model parameters.

**SI 4. Sensitivity analysis of final model**

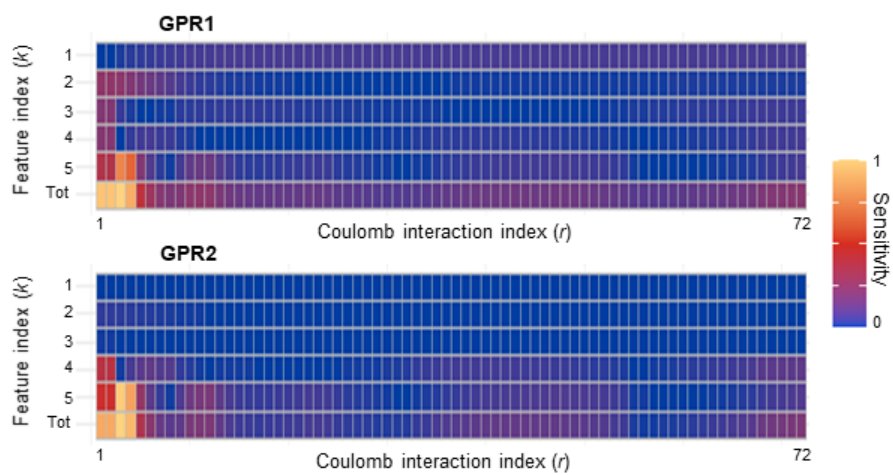

As for Figure 3B (main text), but with sensitivity analysis performed on the final model.

**SI 5. Predictions of final model compared to a single GPR model**

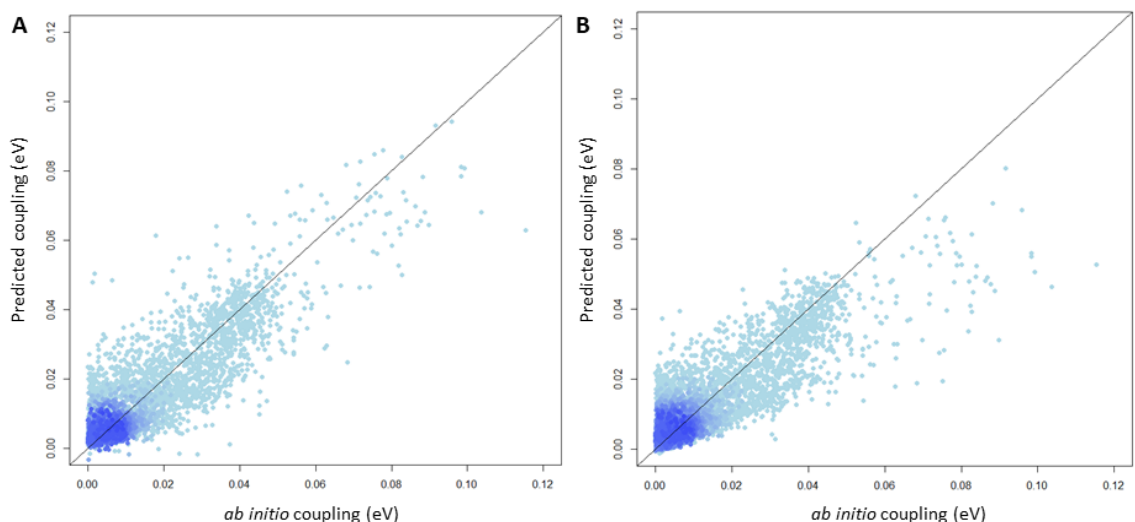

(A) Predicted excitonic coupling energies of the final model compared to *ab initio* values for all extracted pentacene dimers (identical to Figure 4A). (B) Predicted energies of a model consisting of a single Gaussian process regression component. See section 3.3 for details.

For clarity, each point is coloured according to the density of data in its vicinity. Concretely, for each point, the number of other points within a 0.01 eV radius were counted. The 'density' was defined as this number divided by 4927 (the total number of data points).

Linear regression on the data in Figure A yields the result $y = (0.0024 \pm 0.0001) + (0.80 \pm 0.06)x$, where errors refer to one standard error, and $R^2 = 0.77$. For Figure B, we obtain $y = (0.0032 \pm 0.0001) + (0.69 \pm 0.006)x$ and $R^2 = 0.74$.

**SI 6. Predicted diffusion tensor elements**

| | *Ab initio*-calculated couplings | Final model-calculated couplings | Large training set-calculated couplings* |
|---|---|---|---|
| **Diffusion Coefficient** ($\times 10^{-3}$ cm$^2$s$^{-1}$) | 1.630 ± 0.011 | 1.547 ± 0.005 | 1.544 ± 0.006 |
| **Eigenvalues of the Diffusion Tensor ($\times 10^{-3}$ cm$^2$s$^{-1}$)** | | | |
| Major | 1.815 ± 0.014 | 1.686 ± 0.017 | 1.679 ± 0.012 |
| Middle | 1.551 ± 0.016 | 1.492 ± 0.007 | 1.512 ± 0.006 |
| Minor | 1.525 ± 0.012 | 1.462 ± 0.014 | 1.440 ± 0.006 |
| **Ratio (Major: Middle: Minor)** | 1.19: 1.02: 1.00 | 1.15: 1.02: 1.00 | 1.17: 1.05: 1.00 |
| **Diffusion Tensor ($\times 10^{-3}$ cm$^2$s$^{-1}$)** | | | |
| xx | 1.593 ± 0.020 | 1.492 ± 0.012 | 1.500 ± 0.010 |
| yy | 1.682 ± 0.015 | 1.612 ± 0.008 | 1.613 ± 0.012 |
| zz | 1.616 ± 0.007 | 1.536 ± 0.010 | 1.518 ± 0.006 |
| xy | -0.092 ± 0.009 | -0.061 ± 0.012 | -0.098 ± 0.007 |
| yz | -0.108 ± 0.011 | -0.079 ± 0.011 | -0.039 ± 0.010 |
| xz | 0.056 ± 0.012 | 0.033 ± 0.006 | 0.008 ± 0.003 |
| **Ratio (xx: yy: zz: xy: yz: xz)** | 28.5: 30.1: 28.9: 1.65: 1.93: 1.00 | 28.5: 30.8: 29.3: 1.16: 1.51: 0.64 | 28.5: 30.6: 28.8: 1.86: 0.75: 0.15 |

* This model used all dimers to train the SVM component, 2000 dimers to train GPR1, and 2200 dimers to train GPR2